\pdfoutput=1
\documentclass[11pt]{article}
\usepackage{ACL}

\usepackage{times}
\usepackage{latexsym}
\usepackage[T1]{fontenc}
\usepackage[utf8]{inputenc}
\usepackage{microtype}
\usepackage{inconsolata}

\usepackage{hyperref}       
\usepackage{url}            
\usepackage{booktabs}       
\usepackage{amsfonts}       
\usepackage{bbm}
\usepackage{nicefrac}       
\usepackage{xcolor}         
\usepackage{amsmath}
\usepackage{multirow}
\usepackage{array}
\usepackage{float}
\usepackage{threeparttable}
\usepackage{graphicx}
\usepackage{geometry}
\usepackage{quoting}
\usepackage{xspace}

\title{Are Large Language Models (LLMs) Good Social Predictors? }

\date{Nov 2023}

\newcolumntype{L}[1]{>{\raggedright\arraybackslash}p{#1}}
\newcolumntype{C}[1]{>{\centering\arraybackslash}p{#1}}

\newcommand{\FullNameDesign}{\texttt{Social Profile Prediction}\xspace}
\newcommand{\NameDesign}{\texttt{Soc-PRF Prediction}\xspace}

\author{Kaiqi Yang$^1$, Hang Li$^1$, Hongzhi Wen$^1$, Tai-Quan Peng$^2$, Jiliang Tang$^1$, Hui Liu$^1$ \\ $^1$Department of Computer Science and Engineering, Michigan State University \\ $^2$Department of Communication, Michigan State University\\\texttt{kqyang@msu.edu}}

\begin{document}
\maketitle


\begin{abstract}
The prediction has served as a crucial scientific method in modern social studies. With the recent advancement of Large Language Models (LLMs),  efforts have been made to leverage LLMs to predict the human features in social life, such as presidential voting. These works suggest that LLMs are capable of generating human-like responses. However, we find that the promising performance achieved by previous studies is because of the existence of input shortcut features to the response. In fact, by removing these shortcuts, the performance is reduced dramatically. To further revisit the ability of LLMs, we introduce a novel social prediction task, \NameDesign, which utilizes general features as input and simulates real-world social study settings. With the comprehensive investigations on various LLMs, we reveal that LLMs cannot work as expected on social prediction when given general input features without shortcuts. We further investigate possible reasons for this phenomenon that suggest potential ways to enhance LLMs for social prediction. 
\end{abstract}

\section{Introduction~\label{sec:intro}}

Prediction is one of the crucial elements of the scientific methods in social studies~\cite{hofman2017prediction}, with a body of literature~\cite{liben2003link,bakshy2011everyone,cheng2014can} devoted to estimating unobserved or missing data based on observed features. Historically, social prediction is made by statistical models such as linear regression~\cite{uyanik2013study}. With the development of machine learning, supervised methods have been adopted, including random forest, Support Vector Machines (SVM), and neural networks~\cite{chen2021social}. However, the classic machine learning methods notably rely on extensive labeled training datasets, which is labor-intensive, especially in social studies. Additionally, the predictive power of machine learning methods is limited~\cite{mackenzie2015production,athey2018impact} and can hardly model the complicated phenomenon in social life.

Meanwhile, Large Language Models (LLMs) have advanced various text-related tasks, such as question-answering~\cite{zhuang2023toolqa,tan2023can}, code-generation~\cite{nijkamp2022codegen,chen2021evaluating}, and math word problems solving~\cite{zhou2022least,wei2022chain}). The extensive world knowledge~\cite{zhao2023survey} and inference abilities~\cite{creswell2022selection} of LLMs have the potential to mitigate the limitations of classic machine learning methods in predicting features of social datasets. Therefore, there are recent works leveraging LLMs in predicting and simulating human responses, such as voting decisions~\cite{argyle_out_2022,von2023assessing} and political attitude~\cite{rosenbusch2023accurate}. They take advantage of LLMs to augment existing datasets with previously inaccessible features and promising performance is reported. However, our preliminary investigation revisiting the case of voting prediction~\cite{argyle_out_2022} with LLMs indicates that their performance is bolstered by the presence of shortcuts to the desired response features. Specifically, these shortcuts arise when the input contains features directly associated with the feature to be predicted, leading to the exceptional performance of both machine learning models and LLM-based methods. Unfortunately, this efficiency comes with a downside: it overlooks the essential task of uncovering authentic relationships between various features and the label. When these shortcuts are eliminated, we observe a significant decline in the LLMs' effectiveness in addressing social study issues, as detailed in Section \ref{sec:revisit}. This observation leads us to question the true capability of LLMs in social predictions, challenging the prevailing perception of their prowess \cite{argyle2023out}.

To investigate this problem, we introduce a set of studies that utilize general social features as input and simulate real-world settings of feature prediction. In particular, to comprehensively understand the power of LLMs in social predicting, we introduce a new task, \NameDesign (stands for \FullNameDesign). It is designed to predict the social features of individuals while accounting for selected features as explanatory and response variables. Besides, informed by social studies \cite{bailey1998theory}, we categorize social features into two groups that capture individuals' features from different perspectives. This enables us to design three distinct settings, which rigorously categorize features into groups and assess LLMs' predictive capacities. In this work, we evaluate various LLMs, including closed-sourced models GPT 3.5~\cite{OpenAIChatGPT}, GPT 4~\cite{Achiam2023GPT4TR}, and Gemini Pro~\cite{team2023gemini}, as well as lighter open-sourced models like Llama-7B, Llama-7B-chat~\cite{touvron2023llama} and Mistral-7B~\cite{jiang2023mistral}. Our studies suggest that LLMs cannot work on social prediction with general input features without shortcuts. We further explore the potential reasons and future directions to enhance LLMs for social prediction.

\section{Revisit Voting Prediction with LLMs}
\label{sec:revisit}

Large Language Models (LLMs) have demonstrated impressive performance across several societal domains, notably in predicting voting decisions in the United States~\cite{argyle_out_2022,veselovsky2023generating}. In this section, we revisit the voting prediction study with LLMs in \cite{argyle_out_2022}.

The voting prediction study in \cite{argyle_out_2022} adopts the American National Election Studies~\cite{anes} to construct the dataset. ANES is a survey conducted in every presidential election year, with features about American public life, especially political views and decisions. To elicit LLMs' prediction of individual voting decisions, the study selects 10 input features, i.e.,  \texttt{racial/ethnic self-identification}, \texttt{gender}, \texttt{age}, \texttt{ideological self-identification}, \texttt{party identification}, \texttt{political interest}, \texttt{church attendance}, \texttt{if discussing politics with family/friends}, \texttt{patriotism feelings}, \texttt{state of residence}. With all the features above, prompts of individual profiles are constructed with a first-person template and a question to elicit prediction. Then the prompts are fed into LLMs for completion, and the words filled in by LLMs serve as the predicted voting decisions. An example of the prompts is:

\begin{quote}
Racially, I am \underline{white}. I am \underline{male}. Ideologically, I describe myself as \underline{conservative}. Politically, I am a \underline{strong Republican} ... In \underline{2016}, I voted for:
\end{quote}

Intuitively two of the input features are likely equivalent to voting decisions, i.e. \texttt{ideological self-placement} and \texttt{party identification}. It is evident from political studies~\cite{miller1991party,dalton2016party} that given the partisan nature of American politics, voting decisions are closely related to these two features. To validate the intuition, we calculate their {Cramer's V} scores with the voting decisions. {Cramer's V} is a measurement of association between features, ranging from 0 to 1; the score 0 indicates no association and 1 indicates a perfect association. We find that these two features are highly correlated with the vote decisions. For example, in ANES 2016 wave data, these two features have {Cramer's V} scores of $0.86$ and $0.76$, respectively. This indicates their strong associations with voting decisions and consequently, they can become the shortcuts to make predictions.

Next, we conduct further experiments to study the impact of such shortcuts on prediction performance. We choose both GPT-based approaches and classic supervised machine learning models. For GPT-based approaches, we deploy GPT 3.5 as the basis and follow the prompts and zero-shot setting in~\cite{argyle_out_2022} to make predictions based on individual profiles. For classic supervised machine learning models, we choose the Random Forest Classifier. Since the supervised classifier needs labeled data to train, we split the dataset into 80\%/10\%/10\% as training, validation, and test sets. There are two settings for each method: (1) \textbf{Full}, taking the full set of input features; (2) \textbf{w/o shortcut}, taking input features without the shortcut features. To evaluate the performance, we deploy accuracy as the metric, given the balanced distribution of the voting decision (51.9\% vs. 48.1\%). Besides, we adopt Cohen's Kappa $\kappa$ as a metric to evaluate the agreement between prediction and true data. Cohen's Kappa $\kappa$ has values ranging from 0 to 1, where 1 indicates stronger agreement and 0 indicates almost no agreement. 

\begin{figure}[htbp]
    \centering
    \includegraphics[width=0.5\textwidth]{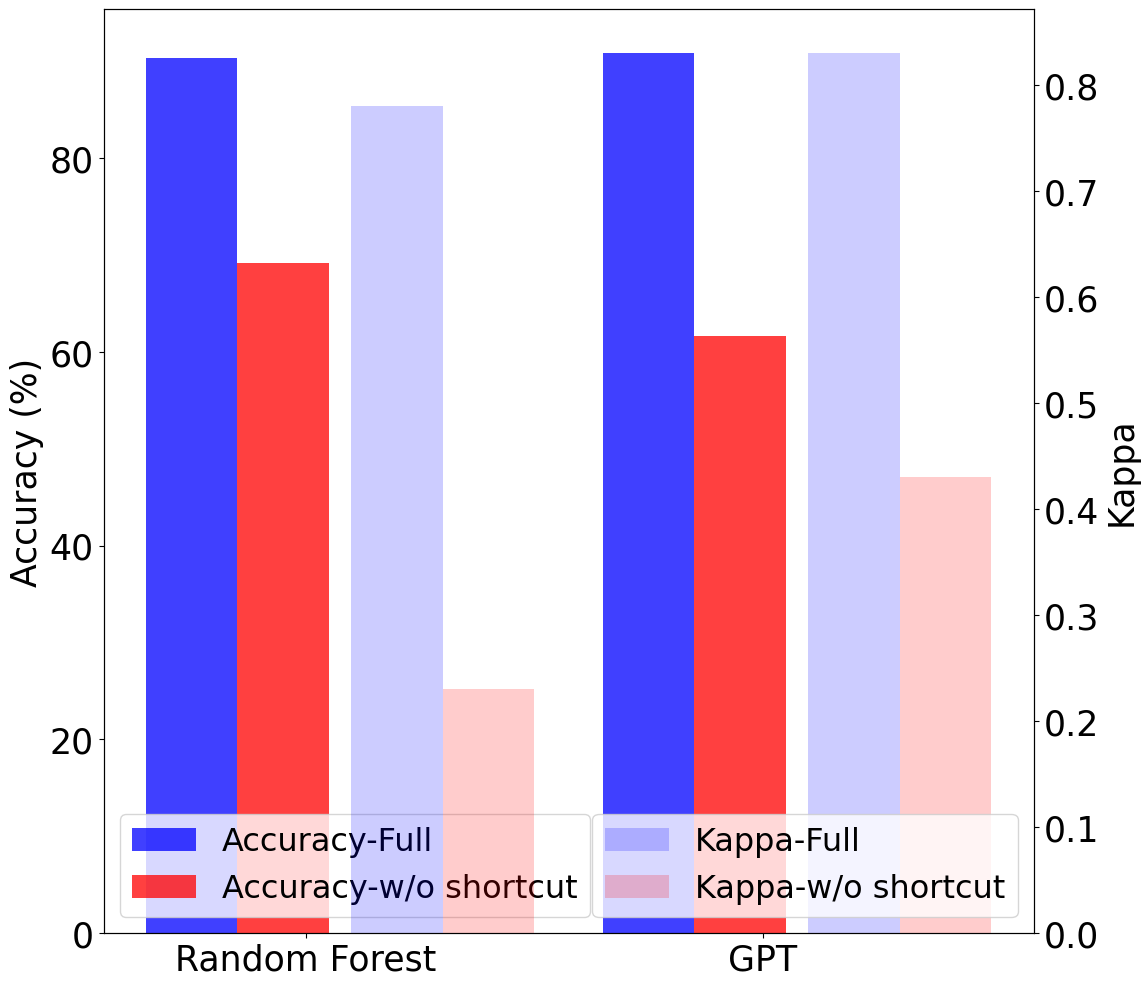}
    \caption{The performance of voting prediction. In this table, GPT stands for the LLM-based approach and we choose GPT 3.5 following ~\cite{argyle_out_2022}. The \textbf{Full} indicates settings with all input features, and \textbf{w/o shortcut} stands for settings without the two shortcut features.}
    \label{fig:voting}
\end{figure}

As shown in Figure~\ref{fig:voting}, the GPT-based approach with all input features achieved the accuracy of {90.82\%} and the {Cohen's Kappa} $\kappa$ of {0.83}, which aligns with~\cite{argyle_out_2022}. However, when removing the two shortcut features, the performance of both methods drops dramatically. For example, the performance of GPT 3.5 drops from the accuracy of 90.82\% and $\kappa$ of 0.83 to an accuracy of 61.60\% and $\kappa$ of 0.43; similarly, the performance of Random Forest drops from 90.29\%, 0.78 to 69.22\%, 0.23. Given the balanced distribution of the voting decision feature, the performance without shortcut features is considerably unsatisfactory. 

Our preliminary study suggests that the promising performance of social prediction via LLMs reported by prior works~\cite{argyle_out_2022} could be from the existence of shortcut features. This finding motivates us to question if LLMs are \textit{really} powerful in social prediction. To explore this question, we design a set of studies that avoid shortcut features as inputs and resemble realistic scenarios in the following section.

\section{Social Profile Prediction}

In this section, we introduce a social prediction task that evaluates the predictive power of LLMs without shortcuts. First, we construct a social prediction dataset based on survey data where we validate that  there are no shortcut features. Then we introduce three settings of evaluations to simulate real-world scenarios. Finally, we show the performance of LLMs' prediction in the proposed settings and discuss the results.

\subsection{Task and Dataset}
As illustrated in Section~\ref{sec:revisit}, the inclusion of shortcut features can affect the evaluation of the power of LLMs in social prediction. Therefore, we design \NameDesign that takes individual social features as input to predict other missing features in profiles. Next, we introduce the dataset for the task.

The dataset derives from Gallup World Poll~\cite {gallup2009world}, one of the most prestigious social surveys that guarantees reliability and offers various types of features. Initialized in 2006, the Gallup World Poll has been conducted in over 150 countries and follows strict random sampling. Questions in the Gallup World Poll are designed by political scientists, measuring key indicators of social life, such as law and order, financial life, civic engagement, etc. Besides, it collects individual demographic data to construct the survey dataset~\cite{tortora2010gallup}. 

In this paper, we construct the dataset based on Gallup World Poll~\cite{gallup2009world} and its corresponding questions. We pick the data from the USA and the data we use in this work is primarily collected between 2016 and 2020. To keep the basic information complete, we remove all the samples with missing data in demographic features. After the data cleaning, the dataset includes 4,941 profiles of American individuals (samples). Additionally, we pick a set of features from the survey to construct the profiles, which encompasses 16 social features reflecting a variety of socio-demographic characteristics, attitudes, and behaviors.

\subsection{Task Settings~\label{sec:setting}}

In social studies, social datasets have predominantly been derived from two methodologies: traditional surveys and online data collection~\cite {couper2017new,diaz2016online,callegaro2014online}. As one characteristic of social features, mutability measures the features' propensity to change or be influenced by social context. Following social studies ~\cite{bailey1992globals,brensinger2021sociology,sen2016race,halley2017sexual}, social features can be roughly divided into two mutability groups: high-mutability and low-mutability. In most social datasets, features with high mutability and low mutability can hardly be collected simultaneously. For example, although survey data collected through in-person interviews is of high quality, it predominantly captures features of low mutability; the dynamic tracking of highly mutable features across all topics and times is nearly impractical due to associated costs. On the other hand, online data collection methods, such as crawling posts from social networking platforms, have the advantage of collecting high-mutability features. Facilitated by natural language processing (NLP) tools~\cite{alghamdi2015survey,vayansky2020review,hussein2018survey,yue2019survey}, real-time human attitudes and opinions are easy to collect. Yet, features of low mutability (e.g. demographic features) often remain inaccessible due to privacy constraints.

Among 16 social features in our proposed dataset, we assign them to low-mutability and high-mutability groups, respectively. The low-mutability features are socio-demographic features, including \texttt{age}, \texttt{gender}, \texttt{marriage}, \texttt{education}, \texttt{employment}, \texttt{income}, and \texttt{urbanicity}. Here the \texttt{age}, \texttt{gender}, \texttt{marriage} status, \texttt{employment} status, \texttt{urbanicity} of residence refer to the individual status when taking the interview, while \texttt{education} refers to the highest completed level of education, and \texttt{income} is the annual household income of last year. The high-mutability features are all about attitudes or behaviors of social life, whose topics include Internet access, social life, economic confidence, civic engagement, and approval of leadership. For each topic, there are one to three questions around it. In save of space, we denote the questions as \texttt{IA}, \texttt{SL1}, \texttt{SL2}, \texttt{EC1}, \texttt{EC2}, \texttt{CE1}, \texttt{CE2}, \texttt{CE3}, \texttt{AL}, respectively. The details of questions are shown in Appendix~\ref{app:dataset}.

According to features' mutability, we design three settings to assess the capability of LLMs in predicting missing features for the individual profiles, which simulate real-world scenarios for social data: from low-mutability features to predicting high-mutability features (for survey data), and from high-mutability features to predicting high-mutability or low-mutability features (for online data). Following the prior works especially~\cite{argyle_out_2022}, we set all three settings as zero-shot, without taking any labeled data as input.

\noindent\textbf{\texttt{low2high}}. In this setting,  the input features possess low mutability, and the output features exhibit high mutability. This setting is designed for traditional survey datasets, where low-mutability features (e.g. demographic features) are comprehensively collected, but high-mutability features (such as attitudes) are sparse. 

To construct prompts of individual profiles, we adopt the template in~\cite{argyle_out_2022}, designing the prompt as a self-description of individuals and inducing the LLMs to predict missing features by completing the self-description. One example of the prompt is:

\begin{quote}
I am a \underline{male} in the USA. I am \underline{42} years old. My current marital status is \underline{married}. My highest completed level of education is \underline{middle} level. My current employment status is \underline{employed}. My Annual Household Income is \$\underline{12600}. I am from \underline{a suburb of a large city}. 

When I'm asked \textit{"Do you have access to the Internet in any way, whether on a mobile phone, a computer, or some other device?"}, my answer is 
\end{quote}

In the provided prompt, the underlined text indicates the values of individual features, and italicized content presents the question to elicit responses. For the subsequent settings, we utilize prompts with the same template.

\noindent\textbf{\texttt{high2low}}: This setting denotes the prediction from high-mutability to low-mutability features. To construct the input profiles of individuals, we take values from all 9 high-mutability features, with a question about one low-mutability feature. Serving as the inverse setting of \texttt{low2high}, this setting is designed for profile construction using online data: aided with NLP tools, the in-time attitudes of individuals can be captured from online posts with ease, yet their demographic features are inaccessible. 

\begin{figure}
    \centering
    \includegraphics[width=0.5\textwidth]{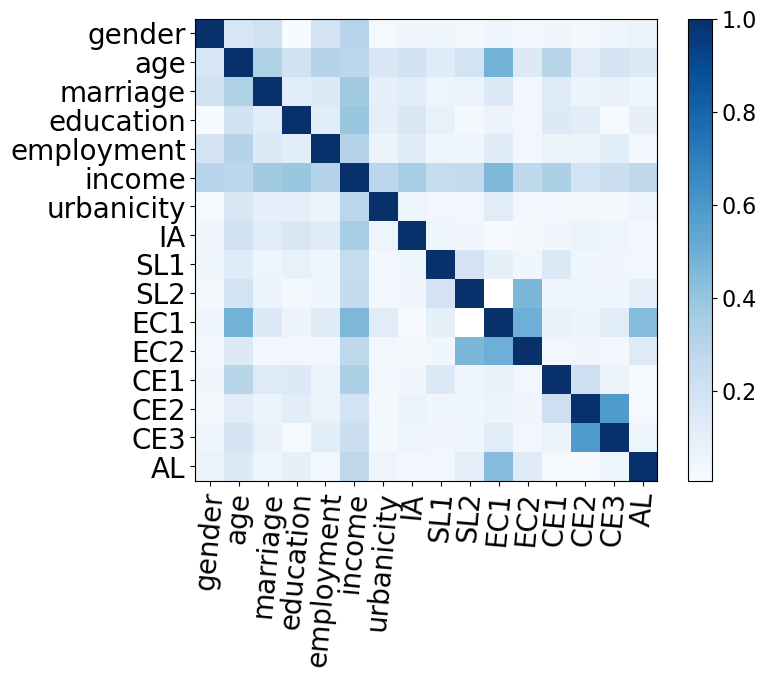}
    \caption{Correlation between features in the dataset. The metric is \textit{Cramer's V}, and values close to 1 indicate strong correlations.}
    \label{fig:corelation}
\end{figure}

\noindent\textbf{\texttt{high2high}}. In this setting, high-mutability features are utilized as input to predict other high-mutability features. Different from \texttt{high2low} setting, in order to avoid shortcuts, the features sharing the same topic with the response feature are excluded from the input profile prompts; rather, we put all high-mutability features of each topic as inputs. This setting simulates a specific real-world scenario, where individuals' attitudes toward one topic are collected, but their opinions on other topics of interest remain unexpressed.

\noindent\textbf{Evaluation Metrics}. Most features in the dataset have imbalanced distributions. For example, the feature \texttt{IA} has 91.82\% samples with "yes" labels, while only 8.18\% samples with "no". In this situation, accuracy is not a suitable metric for imbalanced predictions~\cite{gu2009evaluation}. As a result, we employ AUC as the metric to evaluate the classification performance.

\subsection{Feature Analysis}
To ensure that there are no shortcut features used as input for predictions, we first check the correlations (i.e., the {Cramer's V}) of all feature pairs. As shown in Figure~\ref{fig:corelation}, most of the {Cramer's V} scores are less than 0.5. The maximum (between \texttt{CE2} and \texttt{CE3}) is 0.58, which is merely a moderate level of correlation. This relatively high correlation is because they share a similar topic (i.e., civic engagement). Therefore,  in the following evaluations, we will not consider features that share the same topic with the response as inputs.

\begin{figure*}[th]
    \centering
    \includegraphics[width=0.88\textwidth]{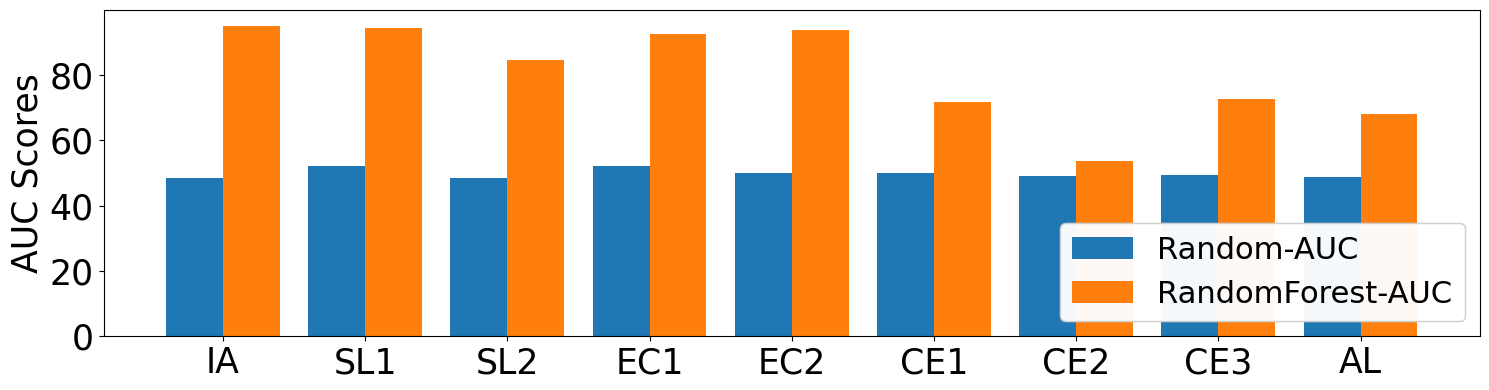}
    \caption{Performance of Random Forest and Random Guessing. The metric is AUC.}
    \label{fig:rf}
\end{figure*}

To evaluate the predictive power of the selected features, we follow the traditional supervised setting. Take the setting of \texttt{low2high} as the example, we first chose Random Forest Classifier as the basis model, and split the dataset by 80\%/10\%/10\% as training, validation, and test sets. The results are shown in Figure~\ref{fig:rf}, the AUC scores are much higher than those of the random guessing method. For example, the AUC score of \texttt{IA} is 95.07, while its corresponding score of random guessing is 48.34.  These observations suggest that though the selected features are not shortcuts, they are still powerful in predicting the target responses.  

\subsection{LLMs as the Predictor}

\begin{figure}
    \centering
    \includegraphics[width=0.51\textwidth]{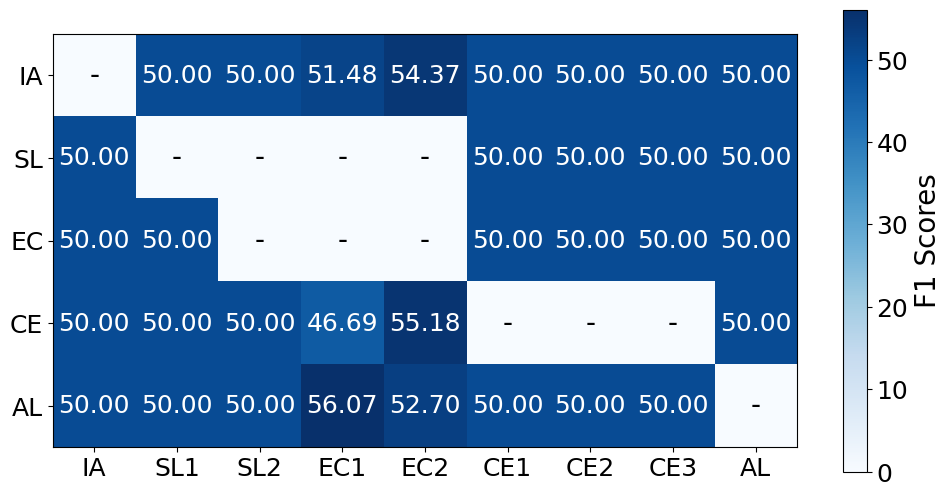}
    \caption{Performance of GPT 3.5 of setting \texttt{high2high}. The metric is \textbf{AUC} score. The sign "-" indicates no valid data, either because the input features (Y-axis) and output features (X-axis) share the same topic, or they are not conducted simultaneously in the survey.}
    \label{fig:s3result}
\end{figure}

In this section, we leverage LLMs for the \NameDesign task with the aforementioned three settings. The results for the three settings are illustrated in Table~\ref{tab:s1full}, Table~\ref{tab:s2}, and Figure~\ref{fig:s3result}, respectively. In the tables, "Random" indicates the random guessing method. Note that for the settings \texttt{high2high}, we only partially show the results because the observations are similar. We note that the performance of LLMs is closely similar to the random guessing and is far from satisfactory. The poor results appear consistently in all the settings and with all the LLMs. These observations indicate that LLMs are struggling to distinguish individual features with the given information in the proposed settings.

\begin{table*}[htbp]
\caption{Performance of LLMs of setting~\texttt{low2high}.}
\centering
\resizebox{0.9\textwidth}{!}{
\begin{tabular}{c|ccccccccc}
\hline
 Model & \texttt{IA} & \texttt{SL1} & \texttt{SL2} & \texttt{EC1} & \texttt{EC2} & \texttt{CE1} & \texttt{CE2} & \texttt{CE3} & \texttt{AL} \\
 \hline

\hline


 Random & 48.34 & 52.09 & 48.47 & 52.12 & 50.07 & 49.89 & 49.16 & 49.32 & 48.60 \\
  \hline
  
 Llama-7B & 50.00 & 50.00 & 50.00 & 48.75 & 55.41 & 50.00 & 50.00 & 50.00 & 50.00  \\
 Llama-7B-chat & 50.00 & 50.00 & 50.00 & 50.95 & 51.80 & 50.00 & 50.00 & 50.00 & 50.00  \\
 Mistral-7B & 50.00 & 50.00 & 50.00 & 53.12 & 56.89 & 50.00 & 50.00 & 50.00 & 50.00 \\
 Gemini Pro & 50.00 & 50.00 & 50.00 & 50.76 & 60.93 & 50.00 & 50.00 & 50.00 & 50.00 \\
 GPT-3.5 & 50.00 & 50.00 & 50.00 & 52.63 & 58.20 & 50.00 & 50.00 & 50.00 & 50.00  \\
 GPT-4 & 50.00 & 50.00 & 50.00 & 53.82 & 56.57 & 50.00 & 50.00 & 50.00 & 50.00 \\
 \hline
\end{tabular}
}
\label{tab:s1full}
\end{table*}
\begin{table*}[htbp]
\caption{Performance of LLMs (GPT 3.5 and GPT 4) of setting~\texttt{high2low}.}
\resizebox{\textwidth}{!}{
\begin{tabular}{c|ccccccc}
\hline
 Model & \texttt{age} & \texttt{gender} & \texttt{marriage} & \texttt{education} & \texttt{employment} & \texttt{income} & \texttt{urbanicity}  \\
 \hline

 Random & 49.50 & 49.62 &49.45 & 49.99 & 50.54 & 48.14 & 50.22\\
 \hline
  Llama-7B & 33.50& 49.81& 50.00& 55.15& 50.00& 50.05& 49.85 \\
  Llama-7B-chat & 40.00&50.00&50.00&35.21&50.33&51.18&50.09 \\
  Mistral-7B & 33.55&49.81&50.00&55.15&50.00&50.05&49.85 \\
  Gemini Pro &38.80&51.14&50.00&66.70&50.00&50.10&49.75\\
  GPT-3.5 & 41.35&50.00&51.29&57.76&49.59&50.95&50.94 \\
  GPT-4 & 40.75&50.00&50.88&65.65&52.01&53.80&52.09 \\
 \hline
\end{tabular}
}
\label{tab:s2}
\end{table*}

\subsection{Discussions}
\subsubsection{Population v.s. Individual}

\begin{figure*}[htbp]
    \centering
    \includegraphics[width=0.95\textwidth]{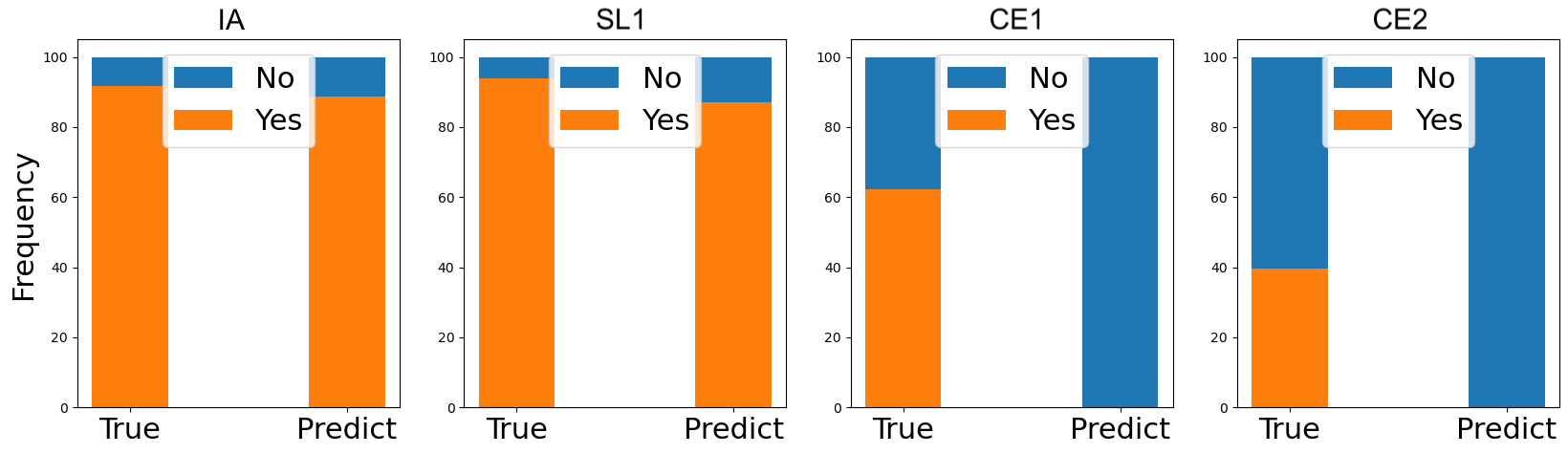}
    \caption{Distributions of Social Features. Note that the last two features (civic engagement behaviors) are more mutable than the first two.}
    \label{fig:analysis1}
\end{figure*}

As shown in the previous subsection, even advanced LLMs like GPT-4 encounter challenges in accurately predicting social features, often yielding outcomes similar to random guessing. To explore the underlying reason for such phenomena, we use the distribution comparisons between predicted features and true ones under the setting \texttt{low2high} as our case studies; the results are shown in Figure~\ref{fig:analysis1}. From the analysis, we have the following observations. First, LLM's prediction of less mutable features, such as IA and SL, is prone to share similar distribution patterns with the true ones. This fact indicates that LLMs do contain some global knowledge about these social features at the population level, but they face challenges in building precise connections to different individual samples during the prediction. Thus, when making individual-level predictions, LLMs may simply generate random samples from the feature distribution. Second, the population-level patterns of highly mutable features, such as different \texttt{CE}s, are seemingly not captured by existing LLMs, and LLMs always prefer to generate negative responses on these features. This fact indicates that building accurate predictors with LLMs for those highly mutable features is more challenging as it not only requires LLMs to establish the connection to individual samples but also external population-level knowledge about the features.

\subsubsection{Incorporating Labeled Data}

Although the performance of zero-shot LLMs' prediction is much worse than our expectation, the strong performance of the random forest classifier in Figure~\ref{fig:rf} indicates that our designed prediction task is reasonable if sufficient labeled data is considered. Based on this finding, we explore the effectiveness of incorporating supervision signals to LLMs based on the  \texttt{low2high} setting as one example. Following the prior studies~\cite{brown2020language,song2023llm}, we leverage the in-context learning ability of LLMs~\cite{dong2022survey,zhang2023trained} to incorporate a few labeled samples as demonstrations. To be specific, for each individual profile, we sample a few other individual profiles from the dataset as a reference, whose \textit{year} and \textit{marriage} features are of the same group. Then, besides constructing vanilla prompts as introduced in Section~\ref{sec:setting}, we add the full information (including the input and output features) of these reference samples in the prompts. Finally, the LLMs are asked to make predictions as the original settings. One example of such prompts is:

\begin{quote}
Here are self-descriptions of \underline{two} people: "I am a \underline{male} in the USA ... my answer is \underline{yes}"; "I am a \underline{female} in the USA ... my answer is \underline{no}"; ...

I am a \underline{male} in the USA. I am \underline{42} years old ...; 

When I'm asked \textit{"Do you have access to the Internet in any way, whether on a mobile phone, a computer, or some other device?"}, my answer is 
\end{quote}

With the augmented prompts, we introduce samples with similar input features and show the true labels of these samples. Like the supervised methods, these demonstrations allow LLMs to make predictions with the help of supervision signals. To provide comprehensive information, the positive and negative labels are balanced within the reference samples. The results of experiments with 2 and 4 demonstrations are shown in Table~\ref{tab:fewshot}. With all selected features, the prompts with demonstrations help LLMs to achieve better prediction performance. However, when incorporating 4 demonstrations into the prompt, we observe marginal or no improvement compared with the setting with 2 demonstrations. This observation suggests that how to effectively incorporate more demonstrations for LLMs for social prediction still faces great challenges.

\subsubsection{Enriching Input Features}

As aforementioned, in reality, no matter survey data or online data, we face constraints to collect sufficient features to describe individuals. In fact, we find that such constraints may also limit the power of LLMs to discriminate different individuals in social prediction. For example, in the \texttt{low2high} setting, we can identify a sub-group from the population characterized by all low-mutability features: \textit{male, married, age between 30 and 60, higher level of education, fully employed, annual household income within the middle 30\%, living in suburbs of large cities}. Within this subgroup, all samples share the same input; however, their responses are varied significantly. For example, to the question \texttt{CE1}, 60.53\% of samples answered "no" and 29.47\% answered "yes". This variability indicates the lack of discriminative features poses significant challenges in precisely predicting individual responses. To unlock the potential of LLMs in predicting social features, enriching input features is crucial.

\begin{table}[htbp]
\centering
\caption{The performance of LLMs (GPT 3.5) with a few demonstrations.}
\resizebox{0.4\textwidth}{!}{
\begin{tabular}{cccc}
\hline
 & \textbf{Zero-Shot} & \textbf{2 Demos} & \textbf{4 Demos} \\
 \hline
\texttt{IA} & 50.00& 71.61 & 82.67 \\
\texttt{SL2} & 50.00& 50.60 & 50.04 \\
\texttt{EC1} & 52.63 & 50.52 & 53.47 \\
\texttt{CE1} & 50.00& 60.17 & 55.34 \\
\texttt{CE2} & 50.00& 53.22 & 52.79 \\
\texttt{AL} & 50.00& 52.03 & 50.80 \\
\hline
\end{tabular}
}
\vspace{-10pt}
\label{tab:fewshot}
\end{table}

\section{Related Work}

In this section, we give an overview of works related to classic machine learning methods and LLMs in social studies. 

\subsection{Machine Learning Methods in Social Prediction}
Quantitative social studies have deployed machine learning methods to model and predict social features in replace of the classical statistical models like OLS and logistic regression~\cite{chen2021social,hindman2015building}. In studies of criminology, K-means and neural networks are deployed to predict criminal behaviors~\cite{reier2020applied}. To predict communication phenomena, the random forest models are utilized to predict the view count of online posts~\cite{hsu2017social}. Besides, \citet{dong2018cnn} uses the SVM classifier to predict public emotions to social news. In addition, deep learning methods have also been widely used in social prediction. For example, CNN models are used to process image data in social studies~\cite{messer2019predicting,dong2018cnn}; RNN and its variants have been utilized to predict sequential data like stock changes~\cite{wu2018hybrid} and user interests~\cite{liu2018social}. However, these methods either suffer from the need for massive training data or the limited predictive power of models, highlighting the need for more powerful prediction models.

\subsection{Using LLMs to Predict Social Features}
With the advent of LLMs, predicting social features with LLMs has been studied by numerous works~\cite{ziems2023can,veselovsky2023generating}. Among social studies, LLMs have been deployed to predict the potential responses or outcomes with ease, especially in scenarios where traditional methods are constrained by cost or ethical concerns. In economics, \citet{phelps2023investigating} studied game theory by examining cooperative and competitive behaviors with LLMs. Within political science, \citet{wu2023large} deployed LLMs to predict the ideological views of politicians. For communication studies, LLMs are used to simulate and predict the potential outcomes of toxic discourse~\cite{tornberg2023simulating}, the political affiliation of Twitter posts~\cite{tornberg2023chatgpt}, etc. 

Additionally, there are growing interests in leveraging LLMs with social survey and interview, aiming to replicate human-like responses to certain questions or attributes of individuals. For example, \citealp{argyle_out_2022} proposed "silicon samples" that deploy LLMs to simulate the people in a survey or interview and predict their partisan views and voting decisions. \citealp{dillion2023can} examined the LLMs response to psychological tests, comparing the decisions and judgements from LLMs and humans. \citealp{aher2023using} proposed sets of  experiments to check LLMs response to interview and games. Besides, fine-tuning LLMs is a promising method for better prediction of social attitudes across years of surveys~\cite{kim2023ai}. At the same time, discussions~\cite{jansen2023employing} are hold about the potential and risks of deploying LLMs in social survey studies.

\section{Conclusion}
In this study, we introduce a survey-based social prediction task to assess the LLMs' predictive ability using general features. Through the replication of experiments and ablation studies of voting prediction tasks, we reveal a significant performance gap between input prompts with and without shortcut features. To further study the LLMs' predictive ability, we propose a real-world survey dataset with rigorously selected features. Based on it, we demonstrate the inability of LLMs to predict social features only with general features. Furthermore, our empirical studies further showcase the potential reasons that constrain the LLMs' predictive power. In our future research, we aim to explore the efficient methods of providing supervision signals and reference information to improve LLMs prediction performance. Moreover, with the abundant social survey and online data, we plan to use fine-tuning methods to fit the LLMs knowledge with social prediction tasks.

\section{Limitations}

In this study, we examine the predictive power of LLMs. We replicate a voting prediction study and find the shortcuts producing plausible results. However, we do not investigate how to select informative features that enhance prediction and what is the upper bound of this task. Then we conduct comprehensive experiments of social prediction, and contend that incorporating labeled data and enrich input features could benefit social prediction. However, we do not provide experiments to validate these suggestions. Lastly, the LLMs are not further tuned and the prompts are adopted from prior works. Tailoring LLMs and proper prompts to meet the needs of social prediction tasks could be a potential direction to explore. Such adjustments could potentially unlock ability and applicability of LLMs in social prediction, further improving the performance of LLMs.


\clearpage
\bibliography{ref}

\begin{thebibliography}{64}
\expandafter\ifx\csname natexlab\endcsname\relax\def\natexlab#1{#1}\fi

\bibitem[{Achiam et~al.(2023)}]{Achiam2023GPT4TR}
Josh Achiam et~al. 2023.
\newblock \href {https://api.semanticscholar.org/CorpusID:257532815} {Gpt-4 technical report}.

\bibitem[{Aher et~al.(2023)Aher, Arriaga, and Kalai}]{aher2023using}
Gati~V Aher, Rosa~I Arriaga, and Adam~Tauman Kalai. 2023.
\newblock Using large language models to simulate multiple humans and replicate human subject studies.
\newblock In \emph{International Conference on Machine Learning}, pages 337--371. PMLR.

\bibitem[{Alghamdi and Alfalqi(2015)}]{alghamdi2015survey}
Rubayyi Alghamdi and Khalid Alfalqi. 2015.
\newblock A survey of topic modeling in text mining.
\newblock \emph{Int. J. Adv. Comput. Sci. Appl.(IJACSA)}, 6(1).

\bibitem[{ANES()}]{anes}
ANES.
\newblock User's guide and codebook for the anes 2012 time series study.

\bibitem[{Anil et~al.(2023)}]{team2023gemini}
Rohan Anil et~al. 2023.
\newblock Gemini: a family of highly capable multimodal models.
\newblock \emph{arXiv preprint arXiv:2312.11805}.

\bibitem[{Argyle et~al.(2022)Argyle, Busby, Fulda, Gubler, Rytting, and Wingate}]{argyle_out_2022}
Lisa~P. Argyle, E.~Busby, Nancy Fulda, Joshua~Ronald Gubler, Christopher~Michael Rytting, and David Wingate. 2022.
\newblock Out of {One}, {Many}: {Using} {Language} {Models} to {Simulate} {Human} {Samples}.
\newblock \emph{Political Analysis}, 31:337 -- 351.

\bibitem[{Argyle et~al.(2023)Argyle, Busby, Fulda, Gubler, Rytting, and Wingate}]{argyle2023out}
Lisa~P Argyle, Ethan~C Busby, Nancy Fulda, Joshua~R Gubler, Christopher Rytting, and David Wingate. 2023.
\newblock Out of one, many: Using language models to simulate human samples.
\newblock \emph{Political Analysis}, 31(3):337--351.

\bibitem[{Athey(2018)}]{athey2018impact}
Susan Athey. 2018.
\newblock The impact of machine learning on economics.
\newblock In \emph{The economics of artificial intelligence: An agenda}, pages 507--547. University of Chicago Press.

\bibitem[{Bailey(1992)}]{bailey1992globals}
Kenneth~D Bailey. 1992.
\newblock Globals, mutables, and immutables: a new look at the micro-macro link.
\newblock \emph{Quality and Quantity}, 26(3):259--276.

\bibitem[{Bailey(1998)}]{bailey1998theory}
Kenneth~D Bailey. 1998.
\newblock A theory of mutable and immutable characteristics: Their impact on allocation and structural positions.
\newblock \emph{Quality and Quantity}, 32(4):383--398.

\bibitem[{Bakshy et~al.(2011)Bakshy, Hofman, Mason, and Watts}]{bakshy2011everyone}
Eytan Bakshy, Jake~M Hofman, Winter~A Mason, and Duncan~J Watts. 2011.
\newblock Everyone's an influencer: quantifying influence on twitter.
\newblock In \emph{Proceedings of the fourth ACM international conference on Web search and data mining}, pages 65--74.

\bibitem[{Brensinger and Eyal(2021)}]{brensinger2021sociology}
Jordan Brensinger and Gil Eyal. 2021.
\newblock The sociology of personal identification.
\newblock \emph{Sociological Theory}, 39(4):265--292.

\bibitem[{Brown et~al.(2020)Brown, Mann, Ryder, Subbiah, Kaplan, Dhariwal, Neelakantan, Shyam, Sastry, Askell et~al.}]{brown2020language}
Tom Brown, Benjamin Mann, Nick Ryder, Melanie Subbiah, Jared~D Kaplan, Prafulla Dhariwal, Arvind Neelakantan, Pranav Shyam, Girish Sastry, Amanda Askell, et~al. 2020.
\newblock Language models are few-shot learners.
\newblock \emph{Advances in neural information processing systems}, 33:1877--1901.

\bibitem[{Callegaro et~al.(2014)Callegaro, Baker, Bethlehem, G{\"o}ritz, Krosnick, and Lavrakas}]{callegaro2014online}
Mario Callegaro, Reginald~P Baker, Jelke Bethlehem, Anja~S G{\"o}ritz, Jon~A Krosnick, and Paul~J Lavrakas. 2014.
\newblock \emph{Online panel research: A data quality perspective}.
\newblock John Wiley \& Sons.

\bibitem[{Chen et~al.(2021{\natexlab{a}})Chen, Tworek, Jun, Yuan, Pinto, Kaplan, Edwards, Burda, Joseph, Brockman et~al.}]{chen2021evaluating}
Mark Chen, Jerry Tworek, Heewoo Jun, Qiming Yuan, Henrique Ponde de~Oliveira Pinto, Jared Kaplan, Harri Edwards, Yuri Burda, Nicholas Joseph, Greg Brockman, et~al. 2021{\natexlab{a}}.
\newblock Evaluating large language models trained on code.
\newblock \emph{arXiv preprint arXiv:2107.03374}.

\bibitem[{Chen et~al.(2021{\natexlab{b}})Chen, Wu, Hu, He, and Ju}]{chen2021social}
Yunsong Chen, Xiaogang Wu, Anning Hu, Guangye He, and Guodong Ju. 2021{\natexlab{b}}.
\newblock Social prediction: a new research paradigm based on machine learning.
\newblock \emph{The Journal of Chinese Sociology}, 8:1--21.

\bibitem[{Cheng et~al.(2014)Cheng, Adamic, Dow, Kleinberg, and Leskovec}]{cheng2014can}
Justin Cheng, Lada Adamic, P~Alex Dow, Jon~Michael Kleinberg, and Jure Leskovec. 2014.
\newblock Can cascades be predicted?
\newblock In \emph{Proceedings of the 23rd international conference on World wide web}, pages 925--936.

\bibitem[{Couper(2017)}]{couper2017new}
Mick~P Couper. 2017.
\newblock New developments in survey data collection.
\newblock \emph{Annual review of sociology}, 43:121--145.

\bibitem[{Creswell et~al.(2022)Creswell, Shanahan, and Higgins}]{creswell2022selection}
Antonia Creswell, Murray Shanahan, and Irina Higgins. 2022.
\newblock Selection-inference: Exploiting large language models for interpretable logical reasoning.
\newblock \emph{arXiv preprint arXiv:2205.09712}.

\bibitem[{Dalton(2016)}]{dalton2016party}
Russell~J Dalton. 2016.
\newblock Party identification and its implications.
\newblock \emph{Oxford research encyclopedia of politics}.

\bibitem[{Diaz et~al.(2016)Diaz, Gamon, Hofman, K{\i}c{\i}man, and Rothschild}]{diaz2016online}
Fernando Diaz, Michael Gamon, Jake~M Hofman, Emre K{\i}c{\i}man, and David Rothschild. 2016.
\newblock Online and social media data as an imperfect continuous panel survey.
\newblock \emph{PloS one}, 11(1):e0145406.

\bibitem[{Dillion et~al.(2023)Dillion, Tandon, Gu, and Gray}]{dillion2023can}
Danica Dillion, Niket Tandon, Yuling Gu, and Kurt Gray. 2023.
\newblock Can ai language models replace human participants?
\newblock \emph{Trends in Cognitive Sciences}.

\bibitem[{Dong et~al.(2022)Dong, Li, Dai, Zheng, Wu, Chang, Sun, Xu, and Sui}]{dong2022survey}
Qingxiu Dong, Lei Li, Damai Dai, Ce~Zheng, Zhiyong Wu, Baobao Chang, Xu~Sun, Jingjing Xu, and Zhifang Sui. 2022.
\newblock A survey for in-context learning.
\newblock \emph{arXiv preprint arXiv:2301.00234}.

\bibitem[{Dong et~al.(2018)Dong, Peng, Li, and Guan}]{dong2018cnn}
Rida Dong, Oinke Peng, Xintong Li, and Xinyu Guan. 2018.
\newblock Cnn-svm with embedded recurrent structure for social emotion prediction.
\newblock In \emph{2018 Chinese Automation Congress (CAC)}, pages 3024--3029. IEEE.

\bibitem[{Gallup(2009)}]{gallup2009world}
G~Gallup. 2009.
\newblock World poll methodology.
\newblock Technical report, Technical Report, Washington, DC.

\bibitem[{Gu et~al.(2009)Gu, Zhu, and Cai}]{gu2009evaluation}
Qiong Gu, Li~Zhu, and Zhihua Cai. 2009.
\newblock Evaluation measures of the classification performance of imbalanced data sets.
\newblock In \emph{Computational Intelligence and Intelligent Systems: 4th International Symposium, ISICA 2009, Huangshi, China, October 23-25, 2009. Proceedings 4}, pages 461--471. Springer.

\bibitem[{Halley(2017)}]{halley2017sexual}
Janet~E Halley. 2017.
\newblock Sexual orientation and the politics of biology: A critique of the argument from immutability.
\newblock In \emph{Sexual orientation and rights}, pages 3--68. Routledge.

\bibitem[{Hindman(2015)}]{hindman2015building}
Matthew Hindman. 2015.
\newblock Building better models: Prediction, replication, and machine learning in the social sciences.
\newblock \emph{The Annals of the American Academy of Political and Social Science}, 659(1):48--62.

\bibitem[{Hofman et~al.(2017)Hofman, Sharma, and Watts}]{hofman2017prediction}
Jake~M Hofman, Amit Sharma, and Duncan~J Watts. 2017.
\newblock Prediction and explanation in social systems.
\newblock \emph{Science}, 355(6324):486--488.

\bibitem[{Hsu et~al.(2017)Hsu, Lee, Lu, Lu, Lai, Huang, Wang, Lin, and Su}]{hsu2017social}
Chih-Chung Hsu, Ying-Chin Lee, Ping-En Lu, Shian-Shin Lu, Hsiao-Ting Lai, Chihg-Chu Huang, Chun Wang, Yang-Jiun Lin, and Weng-Tai Su. 2017.
\newblock Social media prediction based on residual learning and random forest.
\newblock In \emph{Proceedings of the 25th ACM international conference on Multimedia}, pages 1865--1870.

\bibitem[{Hussein(2018)}]{hussein2018survey}
Doaa Mohey El-Din~Mohamed Hussein. 2018.
\newblock A survey on sentiment analysis challenges.
\newblock \emph{Journal of King Saud University-Engineering Sciences}, 30(4):330--338.

\bibitem[{Jansen et~al.(2023)Jansen, Jung, and Salminen}]{jansen2023employing}
Bernard~J Jansen, Soon-gyo Jung, and Joni Salminen. 2023.
\newblock Employing large language models in survey research.
\newblock \emph{Natural Language Processing Journal}.

\bibitem[{Jiang et~al.(2023)Jiang, Sablayrolles, Mensch, Bamford, Chaplot, Casas, Bressand, Lengyel, Lample, Saulnier et~al.}]{jiang2023mistral}
Albert~Q Jiang, Alexandre Sablayrolles, Arthur Mensch, Chris Bamford, Devendra~Singh Chaplot, Diego de~las Casas, Florian Bressand, Gianna Lengyel, Guillaume Lample, Lucile Saulnier, et~al. 2023.
\newblock Mistral 7b.
\newblock \emph{arXiv preprint arXiv:2310.06825}.

\bibitem[{Kim and Lee(2023)}]{kim2023ai}
Junsol Kim and Byungkyu Lee. 2023.
\newblock Ai-augmented surveys: Leveraging large language models for opinion prediction in nationally representative surveys.
\newblock \emph{arXiv preprint arXiv:2305.09620}.

\bibitem[{Liben-Nowell and Kleinberg(2003)}]{liben2003link}
David Liben-Nowell and Jon Kleinberg. 2003.
\newblock The link prediction problem for social networks.
\newblock In \emph{Proceedings of the twelfth international conference on Information and knowledge management}, pages 556--559.

\bibitem[{Liu et~al.(2018)Liu, Xu, Tang, and Crowcroft}]{liu2018social}
Chi~Harold Liu, Jie Xu, Jian Tang, and Jon Crowcroft. 2018.
\newblock Social-aware sequential modeling of user interests: A deep learning approach.
\newblock \emph{IEEE Transactions on Knowledge and Data Engineering}, 31(11):2200--2212.

\bibitem[{Mackenzie(2015)}]{mackenzie2015production}
Adrian Mackenzie. 2015.
\newblock The production of prediction: What does machine learning want?
\newblock \emph{European Journal of Cultural Studies}, 18(4-5):429--445.

\bibitem[{Messer and Fausser(2019)}]{messer2019predicting}
Uwe Messer and Stefan Fausser. 2019.
\newblock Predicting social perception from faces: A deep learning approach.
\newblock \emph{arXiv preprint arXiv:1907.00217}.

\bibitem[{Miller(1991)}]{miller1991party}
Warren~E Miller. 1991.
\newblock Party identification, realignment, and party voting: Back to the basics.
\newblock \emph{American Political Science Review}, 85(2):557--568.

\bibitem[{Nijkamp et~al.(2022)Nijkamp, Pang, Hayashi, Tu, Wang, Zhou, Savarese, and Xiong}]{nijkamp2022codegen}
Erik Nijkamp, Bo~Pang, Hiroaki Hayashi, Lifu Tu, Huan Wang, Yingbo Zhou, Silvio Savarese, and Caiming Xiong. 2022.
\newblock Codegen: An open large language model for code with multi-turn program synthesis.
\newblock \emph{arXiv preprint arXiv:2203.13474}.

\bibitem[{OpenAI(2022)}]{OpenAIChatGPT}
OpenAI. 2022.
\newblock \href {https://openai.com/blog/chatgpt} {Openai chatgpt}.

\bibitem[{Phelps and Russell(2023)}]{phelps2023investigating}
Steve Phelps and Yvan~I Russell. 2023.
\newblock Investigating emergent goal-like behaviour in large language models using experimental economics.
\newblock \emph{arXiv preprint arXiv:2305.07970}.

\bibitem[{Reier~Forradellas et~al.(2020)Reier~Forradellas, N{\'a}{\~n}ez~Alonso, Jorge-Vazquez, and Rodriguez}]{reier2020applied}
Ricardo~Francisco Reier~Forradellas, Sergio~Luis N{\'a}{\~n}ez~Alonso, Javier Jorge-Vazquez, and Marcela~Laura Rodriguez. 2020.
\newblock Applied machine learning in social sciences: neural networks and crime prediction.
\newblock \emph{Social Sciences}, 10(1):4.

\bibitem[{Rosenbusch et~al.(2023)Rosenbusch, Stevenson, and van~der Maas}]{rosenbusch2023accurate}
Hannes Rosenbusch, Claire~E Stevenson, and Han~LJ van~der Maas. 2023.
\newblock How accurate are gpt-3’s hypotheses about social science phenomena?
\newblock \emph{Digital Society}, 2(2):26.

\bibitem[{Sen and Wasow(2016)}]{sen2016race}
Maya Sen and Omar Wasow. 2016.
\newblock Race as a bundle of sticks: Designs that estimate effects of seemingly immutable characteristics.
\newblock \emph{Annual Review of Political Science}, 19:499--522.

\bibitem[{Song et~al.(2023)Song, Wu, Washington, Sadler, Chao, and Su}]{song2023llm}
Chan~Hee Song, Jiaman Wu, Clayton Washington, Brian~M Sadler, Wei-Lun Chao, and Yu~Su. 2023.
\newblock Llm-planner: Few-shot grounded planning for embodied agents with large language models.
\newblock In \emph{Proceedings of the IEEE/CVF International Conference on Computer Vision}, pages 2998--3009.

\bibitem[{Tan et~al.(2023)Tan, Min, Li, Li, Hu, Chen, and Qi}]{tan2023can}
Yiming Tan, Dehai Min, Yu~Li, Wenbo Li, Nan Hu, Yongrui Chen, and Guilin Qi. 2023.
\newblock Can chatgpt replace traditional kbqa models? an in-depth analysis of the question answering performance of the gpt llm family.
\newblock In \emph{International Semantic Web Conference}, pages 348--367. Springer.

\bibitem[{T{\"o}rnberg(2023)}]{tornberg2023chatgpt}
Petter T{\"o}rnberg. 2023.
\newblock Chatgpt-4 outperforms experts and crowd workers in annotating political twitter messages with zero-shot learning.
\newblock \emph{arXiv preprint arXiv:2304.06588}.

\bibitem[{T{\"o}rnberg et~al.(2023)T{\"o}rnberg, Valeeva, Uitermark, and Bail}]{tornberg2023simulating}
Petter T{\"o}rnberg, Diliara Valeeva, Justus Uitermark, and Christopher Bail. 2023.
\newblock Simulating social media using large language models to evaluate alternative news feed algorithms.
\newblock \emph{arXiv preprint arXiv:2310.05984}.

\bibitem[{Tortora et~al.(2010)Tortora, Srinivasan, and Esipova}]{tortora2010gallup}
Robert~D Tortora, Rajesh Srinivasan, and Neli Esipova. 2010.
\newblock The gallup world poll.
\newblock \emph{Survey methods in multinational, multiregional, and multicultural contexts}, pages 535--543.

\bibitem[{Touvron et~al.(2023)}]{touvron2023llama}
Hugo Touvron et~al. 2023.
\newblock Llama 2: Open foundation and fine-tuned chat models.
\newblock \emph{arXiv preprint arXiv:2307.09288}.

\bibitem[{Uyan{\i}k and G{\"u}ler(2013)}]{uyanik2013study}
G{\"u}lden~Kaya Uyan{\i}k and Ne{\c{s}}e G{\"u}ler. 2013.
\newblock A study on multiple linear regression analysis.
\newblock \emph{Procedia-Social and Behavioral Sciences}, 106:234--240.

\bibitem[{Vayansky and Kumar(2020)}]{vayansky2020review}
Ike Vayansky and Sathish~AP Kumar. 2020.
\newblock A review of topic modeling methods.
\newblock \emph{Information Systems}, 94:101582.

\bibitem[{Veselovsky et~al.(2023)Veselovsky, Ribeiro, Arora, Josifoski, Anderson, and West}]{veselovsky2023generating}
Veniamin Veselovsky, Manoel~Horta Ribeiro, Akhil Arora, Martin Josifoski, Ashton Anderson, and Robert West. 2023.
\newblock Generating faithful synthetic data with large language models: A case study in computational social science.
\newblock \emph{arXiv preprint arXiv:2305.15041}.

\bibitem[{von~der Heyde et~al.(2023)von~der Heyde, Haensch, and Wenz}]{von2023assessing}
Leah von~der Heyde, Anna-Carolina Haensch, and Alexander Wenz. 2023.
\newblock Assessing bias in llm-generated synthetic datasets: The case of german voter behavior.
\newblock Technical report, Center for Open Science.

\bibitem[{Wei et~al.(2022)Wei, Wang, Schuurmans, Bosma, Xia, Chi, Le, Zhou et~al.}]{wei2022chain}
Jason Wei, Xuezhi Wang, Dale Schuurmans, Maarten Bosma, Fei Xia, Ed~Chi, Quoc~V Le, Denny Zhou, et~al. 2022.
\newblock Chain-of-thought prompting elicits reasoning in large language models.
\newblock \emph{Advances in Neural Information Processing Systems}, 35:24824--24837.

\bibitem[{Wu et~al.(2018)Wu, Zhang, Shen, and Wang}]{wu2018hybrid}
Huizhe Wu, Wei Zhang, Weiwei Shen, and Jun Wang. 2018.
\newblock Hybrid deep sequential modeling for social text-driven stock prediction.
\newblock In \emph{Proceedings of the 27th ACM international conference on information and knowledge management}, pages 1627--1630.

\bibitem[{Wu et~al.(2023)Wu, Tucker, Nagler, and Messing}]{wu2023large}
Patrick~Y Wu, Joshua~A Tucker, Jonathan Nagler, and Solomon Messing. 2023.
\newblock Large language models can be used to estimate the ideologies of politicians in a zero-shot learning setting.
\newblock \emph{arXiv preprint arXiv:2303.12057}.

\bibitem[{Yue et~al.(2019)Yue, Chen, Li, Zuo, and Yin}]{yue2019survey}
Lin Yue, Weitong Chen, Xue Li, Wanli Zuo, and Minghao Yin. 2019.
\newblock A survey of sentiment analysis in social media.
\newblock \emph{Knowledge and Information Systems}, 60:617--663.

\bibitem[{Zhang et~al.(2023)Zhang, Frei, and Bartlett}]{zhang2023trained}
Ruiqi Zhang, Spencer Frei, and Peter~L Bartlett. 2023.
\newblock Trained transformers learn linear models in-context.
\newblock \emph{arXiv preprint arXiv:2306.09927}.

\bibitem[{Zhao et~al.(2023)Zhao, Zhou, Li, Tang, Wang, Hou, Min, Zhang, Zhang, Dong et~al.}]{zhao2023survey}
Wayne~Xin Zhao, Kun Zhou, Junyi Li, Tianyi Tang, Xiaolei Wang, Yupeng Hou, Yingqian Min, Beichen Zhang, Junjie Zhang, Zican Dong, et~al. 2023.
\newblock A survey of large language models.
\newblock \emph{arXiv preprint arXiv:2303.18223}.

\bibitem[{Zhou et~al.(2022)Zhou, Sch{\"a}rli, Hou, Wei, Scales, Wang, Schuurmans, Cui, Bousquet, Le et~al.}]{zhou2022least}
Denny Zhou, Nathanael Sch{\"a}rli, Le~Hou, Jason Wei, Nathan Scales, Xuezhi Wang, Dale Schuurmans, Claire Cui, Olivier Bousquet, Quoc Le, et~al. 2022.
\newblock Least-to-most prompting enables complex reasoning in large language models.
\newblock \emph{arXiv preprint arXiv:2205.10625}.

\bibitem[{Zhuang et~al.(2023)Zhuang, Yu, Wang, Sun, and Zhang}]{zhuang2023toolqa}
Yuchen Zhuang, Yue Yu, Kuan Wang, Haotian Sun, and Chao Zhang. 2023.
\newblock Toolqa: A dataset for llm question answering with external tools.
\newblock \emph{arXiv preprint arXiv:2306.13304}.

\bibitem[{Ziems et~al.(2023)Ziems, Held, Shaikh, Chen, Zhang, and Yang}]{ziems2023can}
Caleb Ziems, William Held, Omar Shaikh, Jiaao Chen, Zhehao Zhang, and Diyi Yang. 2023.
\newblock Can large language models transform computational social science?
\newblock \emph{arXiv preprint arXiv:2305.03514}.

\end{thebibliography}

\clearpage
\appendix
\section{Appendix}

\subsection{Questions for Features~\label{app:dataset}}
We categorize the selected 16 features into two groups, i.e. high-mutability and low-mutability features. The details of high-mutability features are shown in Table~\ref{tab:mutfeature} and those of low-mutability features are shown in Table~\ref{tab:immutfeature}. The column "Question Abbrev." indicates the abbreviation of the features, which are broadly used in this work. The column "Question Identifiers" indicates the identifier labels of the corresponding questions in the original Gallup survey. 

\begin{table*}[ht]
\resizebox{0.9\textwidth}{!}{
\begin{tabular}{c|c|c|L{0.5\textwidth}|L{0.2\textwidth}}
\hline
\textbf{Topic} & \textbf{\begin{tabular}[c]{@{}l@{}}Question \\ Abbrev.\end{tabular}} & \textbf{\begin{tabular}[c]{@{}l@{}}Question \\ Identifiers\end{tabular}} & \centering\textbf{Question} & \textbf{Options} \\
\hline
Communication Use  & \texttt{IA} & WP16056 & Do you have access to the internet in any way, whether on a mobile phone, a computer, or some other device? & yes, no \\\hline
Social Life  & \texttt{SL1} & WP27 & If you were in trouble, do you have relatives or friends you can count on to help you whenever you need them, or not? & yes, no \\\cline{2-5} 
 & \texttt{SL2} & WP10248 & In the city or area where you live, are you satisfied or dissatisfied with the opportunities to meet people and make friends? & satisfied, dissatisfied \\\hline
Economic Confidence  & \texttt{EC1} & WP148 & Right now, do you think that economic conditions in this country, as a whole, are getting better or getting worse? & better, worse \\\cline{2-5} 
 & \texttt{EC2} & M30 & How would you rate your economic conditions in this country today – as excellent, good, fair, or poor? & excellent, good, fair, poor \\\hline
Civic Engagement  & \texttt{CE1} & WP108 & Have you donated money to a charity in the past month? & yes, no \\\cline{2-5} 
 & \texttt{CE2} & WP109 & Have you volunteered your time to an organization in the past month? & yes, no \\\cline{2-5} 
 & \texttt{CE3} & WP110 & Have you helped a stranger or someone you did not know who needed help? & yes, no \\\hline
Approval of Leadership & \texttt{AL} & WP150 & Do you approve or disapprove of the job performance of the leadership of this country? & approve, disapprove \\ 
\hline
\end{tabular}
}
\caption{Questions and Options of High-mutability Features of Gallup Dataset. }
\label{tab:mutfeature}
\end{table*}

\begin{table*}[ht]
\resizebox{0.9\textwidth}{!}{
\begin{tabular}{C{0.3\textwidth}|c|c|L{0.5\textwidth}}
\hline
\textbf{Immutable Attribute} & \textbf{\begin{tabular}[c]{@{}l@{}}Question \\ Abbrev.\end{tabular}} & \textbf{\begin{tabular}[c]{@{}l@{}}Question \\ Identifiers.\end{tabular}} & \textbf{Options} \\\hline
Age & \texttt{age} & age & - \\\hline
Gender & \texttt{gender} & WP1219 & 1. Man, 2. Woman \\\hline
Marital Status & \texttt{marriage} & WP1223 & 1. Single/Never been married, 2. Married, 3. Separated, 4. Divorced, 5. Widowed, 6. Domestic Partner; \\\hline
Highest Completed Level of Education & \texttt{education} & WP3117 & 1. Completed elementary education or less (up to 8 years of basic education); 2. Secondary - 3 years Tertiary/Secondary education and some education beyond secondary education (9-15 years of education); 3. Completed four years of education beyond high school and/or received a 4-year college degree; \\\hline
Employment Status & \texttt{employment} & EMP\_2010 & 1. Employed full time for an employer, 2. Out of workforce, 3. Employed part time do not want full time, 4. Employed full time for self, 5. Employed part time want full time, 6. Unemployed; \\\hline
Annual Household Income & \texttt{income} & INCOME\_1 & - \\\hline
Living of Urbanicity & \texttt{urbanicity} & WP14 & 1. A suburb of a large city, 2. A small town or village, 3. A large city, 4. A rural area or on a farm;\\
\hline
\end{tabular}
}
\caption{Questions and Options of Low-mutability Features of Gallup Dataset. }
\label{tab:immutfeature}
\end{table*}

\subsubsection{Feature Convert Methods~\label{app:intconvert}}
In the main experiments, there are features of integer or several classes, such as \texttt{income}, \texttt{employment}, etc. We convert them into groups (with the number of groups no larger than four). For \texttt{income}, we calculate the 35\% and 65\% percentiles of the annual household income. Based on them, we categorize \texttt{income} into three classes: lower level, middle level, and higher level. For features with more than 4 classes, we combine similar classes to make the number of classes as 2 or 3.

\end{document}